# Reversible Basal Plane Hydrogenation of Graphene

Sunmin Ryu,[1] Melinda Y. Han,[2] Janina Maultzsch,[3] Tony F. Heinz,[3] Philip Kim,[4] Michael L. Steigerwald[1] and Louis E. Brus[1]*

[1]**Department of Chemistry, Columbia University, New York, NY 10027, USA**
[2]**Department of Applied Physics and Applied Mathematics, Columbia University, New York, NY 10027, USA**
[3]**Departments of Physics and Electrical Engineering, Columbia University, New York, NY 10027, USA**
[4]**Department of Physics, Columbia University, New York, NY 10027, USA**
*E-mail:leb26@columbia.edu

**Abstract**

We report the chemical reaction of single-layer graphene with hydrogen atoms, generated in situ by electron-induced dissociation of hydrogen silsesquioxane (HSQ). Hydrogenation, forming $sp^3$ C-H functionality on the basal plane of graphene, proceeds at a higher rate for single than for double layers, demonstrating the enhanced chemical reactivity of single sheet graphene. The net H atom sticking probability on single layers at 300K is at least 0.03, which exceeds that of double layers by at least a factor of 15. Chemisorbed hydrogen atoms, which give rise to a prominent Raman D band, can be detached by thermal annealing at 100~200 °C. The resulting dehydrogenated graphene is, however, "activated": when photothermally heated it reversibly binds ambient oxygen, leading to hole doping of the graphene. This functionalization of graphene can be exploited to manipulate electronic and charge transport properties of graphene devices.

**Manuscript Text**

Graphene, a single atomic plane of graphite, has outstanding electronic and structural properties, and is a promising candidate for nano-electronic circuits.[1, 2] With the availability of large-area graphene samples, the current top-down fabrication processes based on lithography may be extended to graphene-based devices. Also, thin graphene layers have been actively studied for applications as transparent electrodes and as nanocomposite materials.[3, 4] The materials chemistry of single-sheet graphene has not yet been explored in detail,[5] and such knowledge is essential for its practical use in technology.

Multilayer graphite shows high in-plane strength due to aromatic bond conjugation, and the basal plane is relatively inert chemically. However, single-layer graphene is significantly more reactive with molecular $O_2$ than graphite.[6, 7] This enhanced reactivity may result from the influence of structural distortion. Free-standing graphene at 23 °C shows spontaneous rippling of ~1 nm magnitude and ~10 nm wavelength.[8] In addition, fluctuating patterns of long-short bond alternation around individual C atoms are predicted.[9] Such structural elasticity[10] apparently enables graphene to conform to atomically rough substrates as seen in recent scanning tunneling microscopy studies on silicon dioxide.[11, 12] The local strain built into the rippled, deformed graphene should stabilize transition states of chemical reactions requiring local $sp^3$ hybridization, as shown in the hydrogenation of $C_{60}$ molecules, carbon nanotubes, and graphite.[13] It is well known that the curved surfaces of carbon nanotubes exhibit higher chemical reactivity than the planar sheets of graphene.[14]

In this letter, we use Raman spectroscopy to study the reaction of graphene with hydrogen atoms generated during electron beam initiated cross-linking of a hydrogen silsesquioxane (HSQ) film coated on the graphene sample. Graphene Raman scattering is sensitive to structure and doping,[15] and does not require electrical contacts. We find the reaction to be significantly faster for single-layer graphene than for double layers. Also, we observe that hydrogenated graphene can be restored by thermal annealing. When subsequently photothermally heated in ambient oxygen, the resulting dehydrogenated graphene becomes reversibly hole-doped. Hydrogenated graphene had drawn interest for its predicted magnetism.[16] This particular reaction is also of technical interest since HSQ is a (negative-tone) electron beam resist used in patterning graphene nanoribbons.[17-19] Reactions of the HSQ-derived H atoms with the graphene basal planes should affect the electronic transport behavior of these nanostructured systems.

**RESULTS**

**A. Hydrogenation**

Single or few-layer graphene samples were deposited by mechanically exfoliating kish graphite on Si wafer coated with a 300 nm thick layer of $SiO_2$.[20, 21] The thickness and structural quality[22, 23] of graphene samples were characterized by micro-Raman spectroscopy under ambient condition.[7] Hydrogen atoms were generated in situ by breaking Si-H bonds of HSQ in the course of e-beam lithography:[24] ~30 nm thick films of HSQ (Dow Corning, FOX) were coated on the graphene samples, and irradiated with 30 keV electrons at various doses (0.5~8 mC/cm$^2$). Following development in tetramethylammonium hydroxide solution, some samples were treated with oxygen plasma to etch graphene areas not protected by the cross-linked HSQ film.[17] In one experiment (sample **I**), a Cr/Au electrode was connected to graphene in order to apply a back-gate voltage.

Sample **II** in Fig. 1a shows several graphene pieces without electrode attachment. Squares and rectangles in single-layer (1L) and double-layer (2L) graphene areas are patterned by e-beam lithography. Raman spectra taken before HSQ e-beam patterning (pristine case in Fig. 1c) show no detectible defect-related Raman D band (~1350 cm$^{-1}$), indicating that both the 1L and 2L sheets are initially free of defects. However, e-beam irradiation of 1L graphene covered with HSQ induces a significant D band intensity, as readily seen in the Raman intensity map (Fig. 1b). Calculations of the Raman response indicate that the D band is induced by local basal plane derivatization that creates $sp^3$ distortion.[25] Fig. 1c shows Raman spectra taken at the center of the 1L and 2L squares: It is remarkable that identical e-beam doses generate virtually no D band on 2L graphene, but a very prominent D band on 1L graphene. While the defects are stable in ambient conditions at low laser intensities, intense laser excitation reduces the D band intensity as described below (for estimated temperature rise, see the Supporting Information).

Pristine 1L graphene without coating with an HSQ film showed negligible D band intensity after the same electron irradiation (see the Supporting Information, Fig. S1). Thus, the employed 30 keV electron dose does not directly create D band defects in the graphene lattice, which is consistent with prior studies on other carbon materials.[26, 27] The electron beam causes Si-H bond scission in HSQ, thus initiating cross-linking.[24] We observe that the Si-H Raman band of HSQ[28] at 2265 cm$^{-1}$ decreases by

~90% in intensity following the e-beam irradiation (Fig. 2). Further, the D band intensity grows approximately linearly as e-beam exposure leads to hydrogen depletion in the HSQ, as indicated by the decreasing Si-H bond intensity. 2L graphene is found to be much less reactive. The D band is seen only at very high e-beam dose and has ~60 times lower $I_D/I_G$ (integrated intensity ratio of the D band to the G band) than 1L graphene (Fig. 2c) for the same dose. Given the same density of defects localized on the top graphene layer, the $I_D/I_G$ ratio is expected to be smaller for 2L than 1L samples because of the presence of an intact lower layer for 2L samples. We estimate that the defect density of 1L graphene is still at least 17 times higher than that for 2L materials, based on the work by Z. Ni et al.[29] (see the Supporting Information). H atoms should be the major mobile radical liberated by HSQ cross-linking: the Si-H bond is weaker than the Si-O bonds, and 2~3 bonds need to be broken simultaneously for Si, O, or SiO to be liberated from HSQ molecules.[28] We conclude that we observe the reaction of H atoms with the graphene basal plane.

The G band energy changes slightly following each step of treatment, as shown in Fig. 3. Deposition of HSQ films and hydrogenation ($I_D/I_G$ ~1) decrease the G band energy by < 2 cm$^{-1}$. $O_2$ plasma treatment leads to an increase < 1 cm$^{-1}$. Physical contact of the ~30 nm HSQ film and subsequent solvent drying might lead to in-plane stress,[29] thus affecting the G band energy. However, the reciprocal relation between the G band energy and its width (Fig. 3) suggests that overall change is mainly driven by chemical doping: as demonstrated by electrical gating experiments the concurrent upshift (downshift) and band narrowing (broadening) of the G band is explained by charge doping.[30, 31] Due to non-adiabatic electron-phonon coupling, the G band shifts upward as the Fermi level is displaced from its neutrality point by doping. At the same time, the width of the G band decreases since Landau damping of the G phonon is not possible when the Fermi level is shifted by more than half of the G band energy from the neutrality point. When hydrogenated, graphene is expected to be electron doped since carbon is slightly more electronegative than hydrogen. This conclusion agrees with a recent DFT calculation[32] on the presence of a hydrogenation-induced Fermi level shift. Electron doping is also consistent with the observed red-shift of the G band upon hydrogenation, since initial graphene on silicon dioxide is lightly hole-doped from the environment.[20]

The reaction of H atoms with multilayer graphite has previously been studied in detail.[33-35] The binding energy of a single H atom is low (~0.7 eV). Isolated adsorbed H atoms ($H_{ad}$) show a relatively small activation energy for desorption (~0.9 eV) and are not stable at room temperature.[33-35] Para or ortho $H_{ad}$ pairs in one benzene ring form at higher coverages. Pairs are more strongly bound and have a significantly larger activation energy for recombinative desorption as $H_2$.[35] We do not know if we are observing single or paired $H_{ad}$ on our graphene samples.

On graphite, H adsorption is reversible: thermal desorption spectra have one major desorption peak at ~200 °C and a minor one at ~290 °C.[36] In a combined STM study, it has been shown that annealing at 423 °C completely removes adsorbed H atoms, restoring the original crystalline lattice without vacancies. In Fig. 4, oven annealing of hydrogenated 1L graphene also shows partial reversibility. After annealing, the Raman D band decreases drastically in intensity relative to the G band. The $I_D/I_G$ ratio decreases by more than a factor of 6 when annealed in Ar at 200 °C (inset of Fig. 4). This comparison also supports our assignment of the D band to the influence of hydrogenation.

Fig. 4 shows that the defects induced by hydrogenation can be thermally healed to a significant extent, largely restoring the original graphene lattice. Intense focused laser radiation also induces a similar effect as shown in Fig. 5c: During repeated Raman measurements, the $I_D/I_G$ ratio gradually decreases because of photothermal desorption of H. The temperature of graphene induced by photothermal heating in Fig. 5 is estimated to be ~60 °C based on the temperature coefficient of the G band energy[37] (see the Supporting Information). The controllable hydrogenation with the high spatial resolution of e-beam lithography and reversible dehydrogenation may find application in patterning graphene into semiconducting or insulating nano-domains. This approach could also be applied to carbon nanotubes and other graphitic materials.

B. **Activation for subsequent reaction**

After thermal annealing, the Raman spectrum of the dehydrogenated graphene is similar to the initial Raman spectrum, except for a residual D band. Nevertheless, after oven annealing the sample is "activated": it shows enhanced chemical doping when photothermally heated in an oxygen atmosphere. In Fig. 5a, the G band energy is plotted as a function of photoirradiation time. While non-oven annealed graphene (hydrogenated or non-hydrogenated) shows less than a 0.5 cm$^{-1}$ increase, the graphene G band blue-shifts by 1.5 cm$^{-1}$ after oven annealed in air or Ar at 100 °C. The simultaneous G band narrowing in Fig. 5b suggests that charge doping occurs. Notably, the blue-shift is mostly attributed to chemical doping caused by molecular oxygen: the G band energy of the photothermally heated graphene decreases (increases) in Ar ($O_2$) atmosphere with concurrent line width broadening (narrowing) (see the Supporting Information, Fig. S2 and S3).

To determine the polarity of charge doped by $O_2$, we titrated with electrically-induced charge. Graphene connected to an external electrode forms a capacitor with the Si back gate; it accumulates electrons (holes) by applying positive (negative) gate voltage.[30, 31] Figures 6a and 6b present the G band energy and line width, respectively, measured for the dehydrogenated graphene (as in Fig. 3) as a function of gate voltage ($V_G$) in both Ar and $O_2$ flow. (For spectra, see the Supporting Information, Fig. S4). In Ar, the G band energy has a minimum at $V_G \approx +5$ V, which indicates that the annealed graphene is barely doped in the Ar environment. In $O_2$, however, the G band energy minimum is located at $V_G > +50$ V, which indicates that the annealed graphene is heavily doped with holes (> 4X10$^{12}$ holes/cm$^2$), requiring additional electrons for compensation. The line width shown in Fig. 6b correlates well with the G band energy, as discussed earlier. Thus, we conclude that molecular oxygen binds reversibly to oven-annealed "activated" graphene, leading to hole doping at room temperature.

**DISCUSSION**

A. **Hydrogenation**

Near room temperature, H atoms are more reactive with 1L than 2L graphene. The enhanced reactivity of single-layer graphene was also seen earlier in oxidative etching at higher temperatures.[7] This suggests that 1L graphene shows a distortion, or degree of freedom, not present in 2L. 1L graphene does not have multilayer π-stacking, which favors a flat structure and is known to increase reaction activation barriers.[6] 1L graphene has significant ripples both when free-standing[8] and when supported on $SiO_2$ substrates.[11, 12, 38] The ripples were found to decrease with increasing thickness of the graphene samples.[8, 38] Such out-of-plane ripples induce some $sp^3$ hybridization in otherwise $sp^2$-hybridized carbons. A recent Monte Carlo simulation confirming energetically stable ripples in free standing graphene also concluded that the C-C bond length of rippled graphene has a significantly broader variation than that of flat graphene.[9] These factors naturally explain the enhanced reactivity of 1L graphene.

Although the density of the defects has not been determined experimentally, an order-of-magnitude estimation can be made based on simulations of the Raman spectra of hydroxylated graphene.[25] For evenly distributed 1,2-hydroxyl pairs with an OH density of $4.8 \times 10^{14}$ /cm$^2$ (0.13 ML), the D band is predicted to be as intense as G band. Because the $I_D/I_G$ ratio of the hydrogenated graphene in Fig. 2a (2 mC/cm$^2$) is roughly unity, the H defect number density can be estimated to be $\sim 5 \times 10^{14}$ /cm$^2$. Given the average HSQ film thickness of 30 nm, density[39] of $\sim$1.3 g/cm$^3$, and molecular weight of 424 as $H_8Si_8O_{12}$, the maximum integrated H atom flux available is $4.4 \times 10^{16}$ /cm$^2$ (equivalent to $\sim$10 ML). This gives $\sim$0.03 for net sticking probability ($S$) of H atom on 1L graphene at 300K, assuming that $\sim$70% of Si-H bonds are dissociated (Fig. 2c) and that half of the liberated H atoms reach graphene surface. We neglect here H recombination or other loss mechanisms; such processes would make $S$ higher.

Based on the G band energy change, the amount of charge transferred from a single hydrogen atom can be estimated. The G band energy shift of $\leq 2$ cm$^{-1}$ (Fig. 3) caused by the hydrogenation giving $I_D/I_G \sim 1$ (Fig. S1b) corresponds roughly to a change in charge density of $\sim 1.3 \times 10^{12}$ $e$/cm$^2$.[30] Assuming the number density of H defects is $\sim 5 \times 10^{14}$ /cm$^2$ (0.13 ML), $\sim$0.003$e$ charge is donated by each chemisorbed H atom. Compared to chemical doping by weakly interacting molecules such as water, $NH_3$, CO, and NO,[40] the estimated charge transfer is several times lower. This appears reasonable considering the largely non-polar nature of C-H bonds. Controllable chemical doping by forming covalent bonds of varying polarity can be exploited to modify the electronic properties of graphene for possible device applications.

B. **Activation for subsequent reaction**

We find that graphene oven annealed at 100 °C is more reactive than pristine graphene. Recently, Li et al. observed that graphene annealed at >250 °C is "activated" and, unlike the initial pristine graphene, reversibly binds oxygen molecules under ambient. We infer that thermal or photothermal annealing generates at present unknown structural changes. Molecular oxygen in its excited singlet state is known to form endoperoxides with hundreds of aromatic compounds.[41] Due to the $sp^3$ character of carbon atoms connected to $O_2$ in endoperoxides, strained π-systems have much higher affinity to $O_2$ than do planar ones. Owing to its severe strain as well as extended conjugation, helianthrene is known to bind even ground state triplet $O_2$ to form endoperoxides.[42] Carbon nanotubes, systematically strained by curvature yet showing no D band, also bind $O_2$ to form endoperoxides.[43]

Thermal annealing may remove water and extraneous organic matter initially present between graphene layers and the substrate $SiO_2$. Graphene annealed in direct contact with atomically rough $SiO_2$ may further deform. Graphene on $SiO_2$ is 60% smoother in height variation than the bare $SiO_2$ surface,[12] which implies that the annealed graphene sheet has deformed in response to the substrate.[7] The substrate-mediated thermal activation is thought to generate $O_2$-binding sites with a high degree of out-of-plane deformation. Compared to graphene annealed at >250 °C, the doping level of the photothermally heated graphene is much lower, which implies that further activated deformation can occur. We doubt that activation represents vacancy formation: oxygen would bind strongly and irreversibly to a vacancy cite.[44]

In conclusion, we have shown that 1L graphene can be more easily hydrogenated than 2L graphene near room temperature. This enhanced reactivity is attributed to the lack of π-stacking and/or out-of-plane deformation needed to stabilize the transition state of the hydrogenation reaction. The hydrogenated graphene can be restored by thermally desorbing bound hydrogen atoms. Dehydrogenated graphene on $SiO_2$ is "activated" and exhibits enhanced chemical doping caused by oxygen molecules when photothermally heated. The bound oxygen molecules lead to reversible hole doping of graphene.

**Acknowledgment.** This work was funded by the Department of Energy under Grant Nos. DE-FG02-03ER15463 (T.F.H) and DE-FG02-98ER14861 (L.E.B), and by National Science Foundation under No. DMR-0349232 (P.K.). We acknowledge financial support from the Nanoscale Science and Engineering Initiative of the National Science Foundation under NSF Award No. CHE-06-41523 by the Nanoelectronics Research Initiative (NRI) of the Semiconductor Research Corporation, and by the New York State Office of Science, Technology, and Academic Research (NYSTAR). J.M. acknowledges support from the Alexander von Humboldt foundation.

**Supporting Information Available:** The dependence of the D band sensitivity on graphene thickness, photothermally induced reversible $O_2$ binding on pre-annealed graphene, temperature of graphene under photothermal heating, effects of e-beam irradiation on the Raman spectra of bare graphene compared to HSQ-covered graphene, raw Raman scattering data for Fig. 6.

**Figures and Captions**

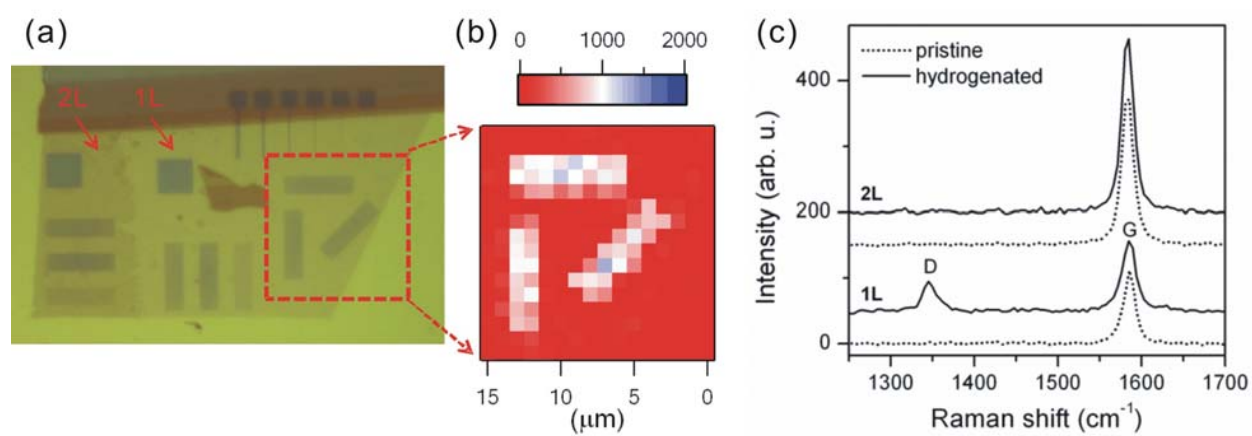

**Figure 1.** (a) Optical micrograph of e-beam patterned sample **II**, which contains 1L, 2L and thick sheets of graphene. The squares and rectangles are crosslinked HSQ etch masks. Non-crosslinked HSQ has been removed by the developer. The 1L area in the dashed square is 15X15 μm$^2$ in size. (b) The D band intensity Raman map for the dashed square in (a). (c) Raman spectra taken at the center of 1L and 2L graphene squares (area: 4X4 μm$^2$) shown in (a), before (dotted) and after (solid, displaced for clarity) hydrogenation. Data in (b) and (c) were obtained in ambient conditions with $\lambda_{exc}$ = 514.5 nm. 3 mW power was focused onto a spot of ~1 μm in diameter. The integration time for each pixel was 20 sec.

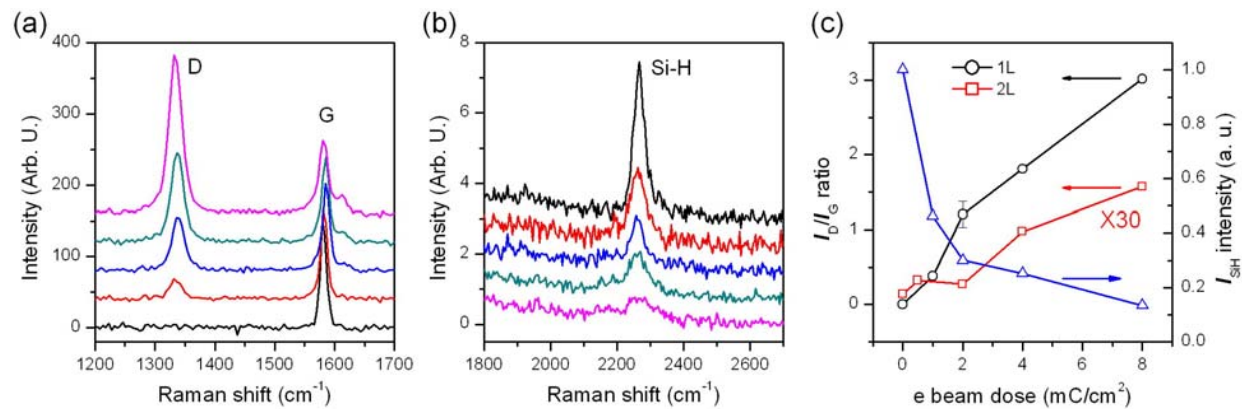

**Figure 2.** (a) Raman spectra of 1L graphene sheets (sample **V**), at various stages of hydrogenation: the e-beam dose was 0, 1, 2, 4, and 8 mC/cm$^2$, respectively, from the bottom spectrum to the top one (displaced for clarity). (b) Raman spectra of HSQ films on SiO$_2$, irradiated with the same e-beam: the dose was 0, 1, 2, 4, and 8 mC/cm$^2$, respectively, from the top spectrum (displaced for clarity) to the bottom one. Each spectrum was obtained from a region of the HSQ film located within 10 μm from each of the above graphene sheets. (c) Integrated intensity ratio of the D band to G band ($I_D/I_G$) as a function of e-beam dose: 1L (circles) and 2L (squares) graphene. Triangles represent the Si-H band intensity as a function of e-beam dosage. All spectra were obtained in ambient conditions with the 514.5 nm excitation laser focused to a spot size of ~1 μm diameter. The laser power was 0.1 and 3 mW for (a) and (b), respectively.

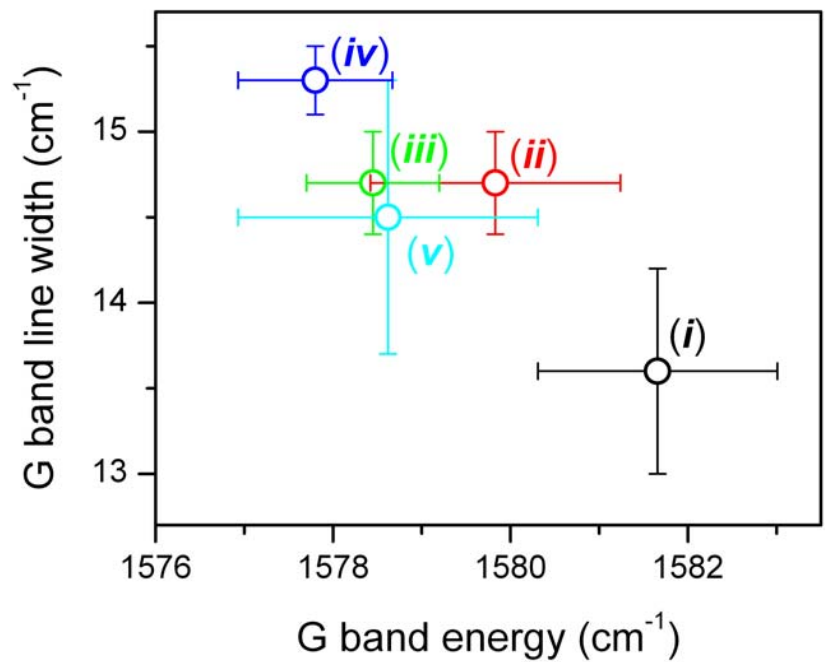

**Figure 3.** G band line width vs energy, obtained for sample **III** following each treatment step: (*i*) no treatment-pristine graphene, (*ii*) after HSQ film deposition, (*iii*) after hydrogenation (e-beam dose: 0.5 mC/cm$^2$), (*iv*) after additional hydrogenation (accumulated e-beam dose: 1.0 mC/cm$^2$), (*v*) after oxygen plasma treatment. The error bars represent standard deviations for 10 spots measured in one graphene sample. All data were obtained in ambient conditions. The excitation laser operated at a wavelength of $\lambda_{exc}$ = 514.5 nm and a power of 3 mW was used in a spot size of ~1 μm diameter. The line width includes an instrument response function of 6.0 cm$^{-1}$.

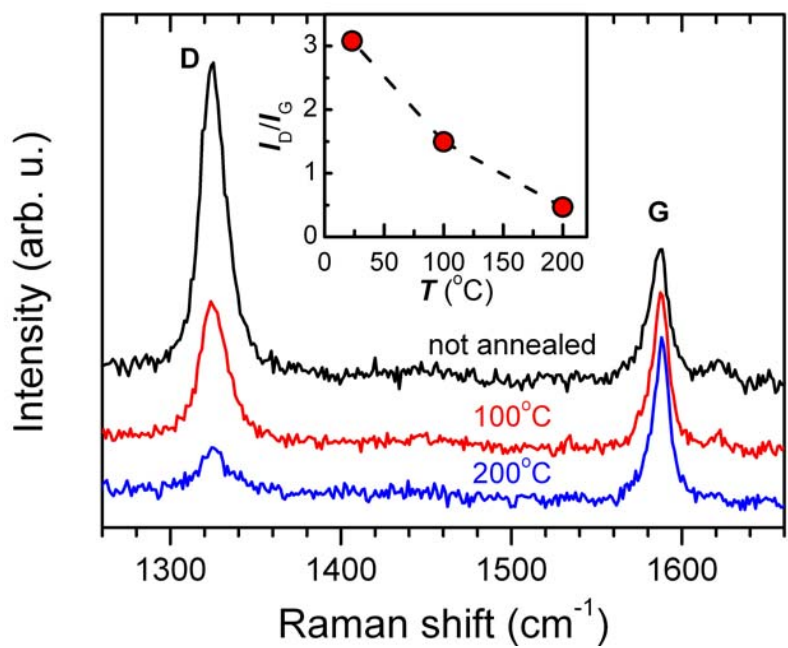

**Figure 4.** Raman spectra of hydrogenated 1L graphene (sample **I**), taken before and after oven annealing for 1 hour at 100 °C (in air) and 200 °C (in Ar). The inset shows the ratio of $I_D/I_G$ as a function of the annealing temperature. The spectra were obtained under ambient conditions. The Raman pump laser wavelength operated at a wavelength of $\lambda_{exc}$ = 632.8 nm, with a power of 4 mW focused to a spot of ~ 2 μm diameter.

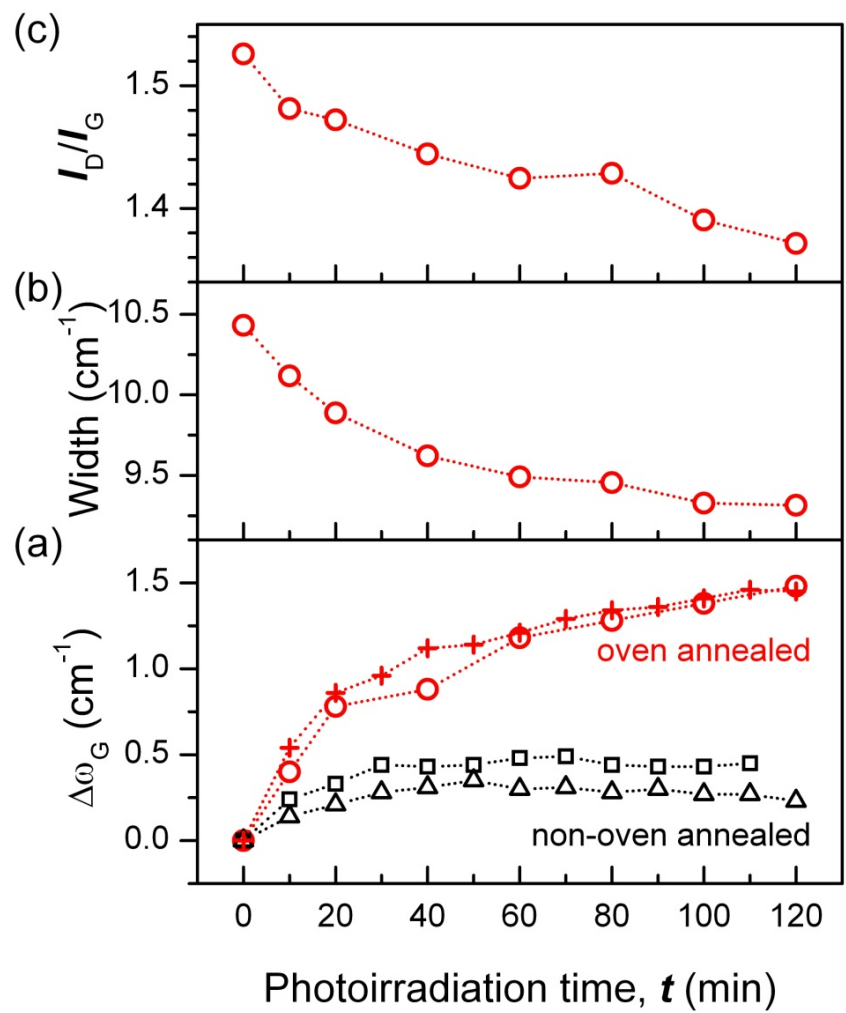

**Figure 5.** Effects of prolonged photoirradiation in $O_2$ atmosphere on 1L graphene Raman spectra: (a) G band energy change, $\Delta\omega_G = \omega_G(t) - \omega_G(t = 0)$, (b) G band line width, and (c) $I_D/I_G$ as a function of photoirradiation time ($t$). The history of the samples is as follows. Circles (sample **I**) hydrogenated (0.5 mC/cm$^2$) followed by oven annealing in air at 100 °C; crosses (sample **VI**) hydrogenated (0.5 mC/cm$^2$) followed by oven annealing in Ar at 100 °C; triangles (sample **VI**) hydrogenated (0.5 mC/cm$^2$) but not oven annealed; squares (sample **V**) HSQ-coated but non-hydrogenated and non-oven annealed. 2 μm spots of each sample were continuously irradiated with a 632.8 nm laser (4 mW) for 120 minutes during the course of the Raman measurements. The line width includes an instrument response function of 3.7 cm$^{-1}$.

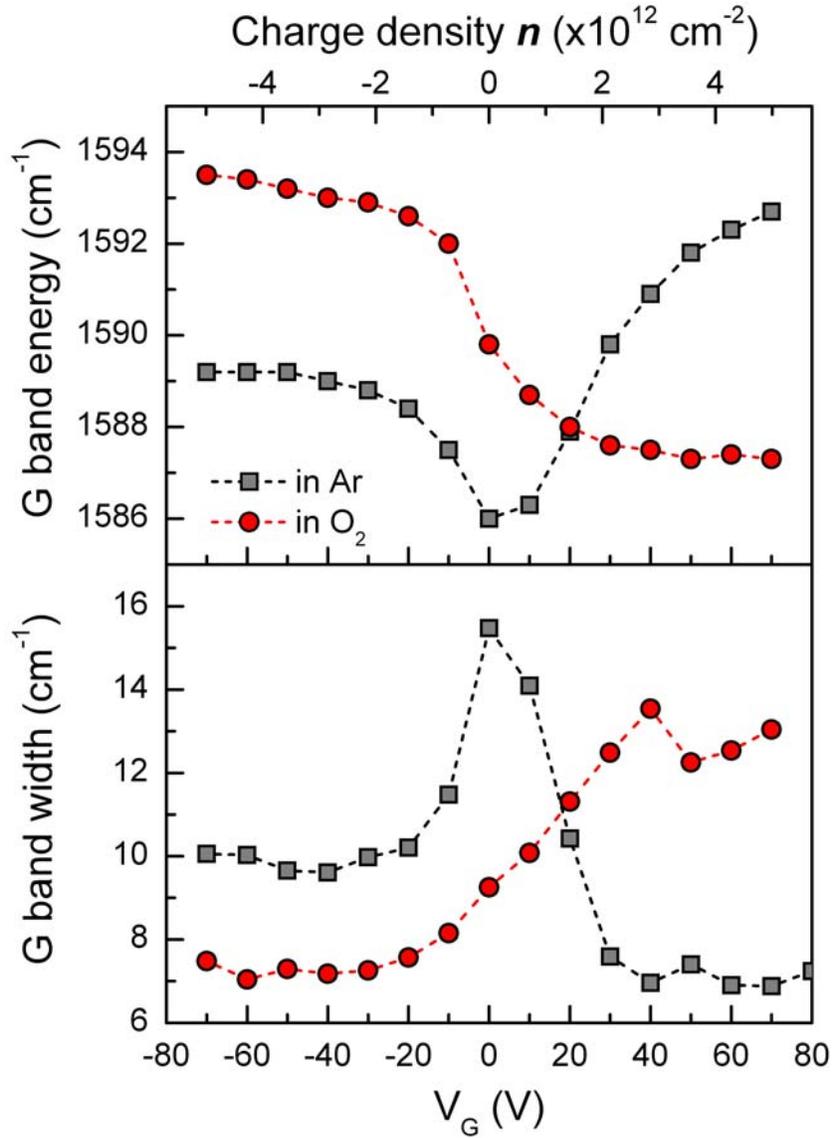

**Figure 6.** G band energy (upper) and line width (lower) of 1L graphene (sample **I**), obtained as a function of back-gate voltage ($V_G$) in Ar and $O_2$. The graphene sample was dehydrogenated at 100 °C in air in the oven, followed by further photothermal heating. The charge density (***n***) of upper x-axis refers to the number density of electrons in graphene induced by the electrical gating. According to Ref. 30, ***n*** = $C_G V_G/e$, where $C_G$ is 115 aF/μm². The line width includes an instrument response function of 3.7 cm$^{-1}$.

**TOC Graphic**

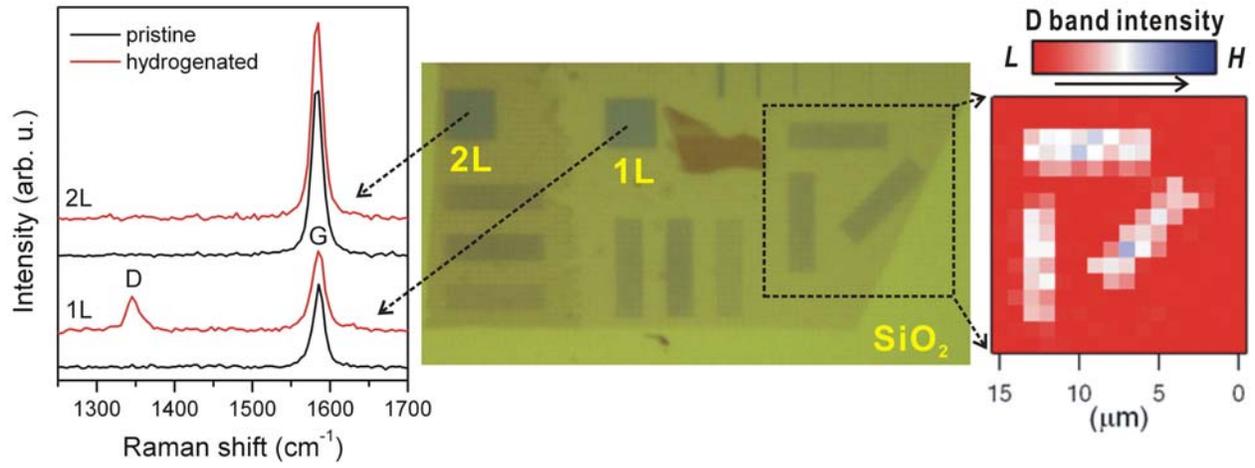

# Supporting Information for

# Reversible Basal Plane Hydrogenation of Graphene

Sunmin Ryu,[1] Melinda Y. Han,[2] Janina Maultzsch,[3] Tony F. Heinz,[3] Philip Kim,[4] Michael L. Steigerwald[1] and Louis E. Brus[1]*

### A. Dependence of the Raman D band sensitivity on the thickness of the graphene sample

Graphite or multilayer graphene with defects localized only on the top layer is lower in the D band to G band intensity ratio ($I_D/I_G$) than 1L graphene with the same defect density because Raman scattering also probes tens of intact sub-layers for multilayers. However, it has not been quantitatively determined how much the $I_D/I_G$ will be attenuated as increasing thickness. The attenuation factor in the $I_D/I_G$ for 2L graphene can be estimated based on a recent paper[S1] which reported that pulsed laser deposition of $SiO_2$ generates defects in 1L and multilayer graphene: 2L graphene is ~3.5 times lower in the $I_D/I_G$ than 1L graphene. Assuming the top layer of 2L graphene has a defect density equal to that of the 1L sample, the attenuation factor should be ~3.5; because 2L graphene is expected to be less reactive (at most equally) to general chemical attacks than 1L graphene,[S2] the estimated value is an upper limit of the attenuation factor. Since the $I_D/I_G$ of the hydrogenated 2L graphene is ~60 times smaller than that of 1L graphene (Fig. 2c), the defect density of 2L graphene is at least 17 times lower than that of 1L graphene assuming that the $I_D/I_G$ is proportional to the defect density. (It would be lower still, if the $I_D/I_G$ scales with spacing between neighboring defects) In other words, hydrogenation proceeds at a 17 times slower rate for 2L graphene than 1L graphene.

### B. Photothermally induced reversible $O_2$ binding on pre-annealed graphene

Thermally annealed (dehydrogenated) graphene is more reactive with molecular oxygen than non-annealed graphene: the doping level represented by the G band energy increases during photoirradiation as shown in Fig. 5a. Sample **I** was partly dehydrogenated by annealing at 100 °C, and then studied in a flowing gas optical chamber at 23 °C. All the data were taken at one spot which was further annealed photothermally to saturate the change in the G band energy during Raman measurements (see Fig. 5a). In Fig. S2a, the sample is exposed to flowing Ar for several hours, and then switched to flowing $O_2$. As time evolves (bottom spectrum to top spectrum), the G band upshifts and narrows. In Fig. S2b, the opposite behavior in the G band energy and width can be seen when $O_2$ flow is replaced by Ar flow. Detailed peak positions and line widths are shown in Fig. S3. The current observation indicates that the photothermally annealed graphene reversibly binds molecular oxygen, which electrically dopes graphene. This observation also highlights the fact that moderate annealing (~100 °C) affects chemical reactivity of graphene. It is notable that the line width of the D band does not respond to oxygen while its peak position varies reversibly as does that of the G band (see Fig. S3).

### C. Temperature of graphene under photothermal heating

Despite exposure to intense excitation by the Raman pump laser (e.g., an irradiance of 130 kW/cm$^2$ associated with 4 mW of laser power focused to a 2 μm spot), the inferred temperature rise in the graphene sample was modest. The laser-induced temperature rise ($\Delta T$) was measured by monitoring the G band energy, which decreases at a rate of -0.016 cm$^{-1}$/°C.[S3] For a typical irradiance of 130 kW/cm$^2$, the HSQ-covered graphene (sample **I** in Fig. 5) exhibited a red-shift of the G band of 0.6 cm$^{-1}$ or $\Delta T$ of 37 °C. To saturate the photoinduced change in the G band energy shown in Fig. 5, one particular area in sample **I** was exposed to extended photoirradiation (4 mW) before the temperature measurement. Thus, the effective temperature ($T$) of graphene during the photothermal heating in Fig. 5 is ~60 °C.

**Supporting Figures and Captions**

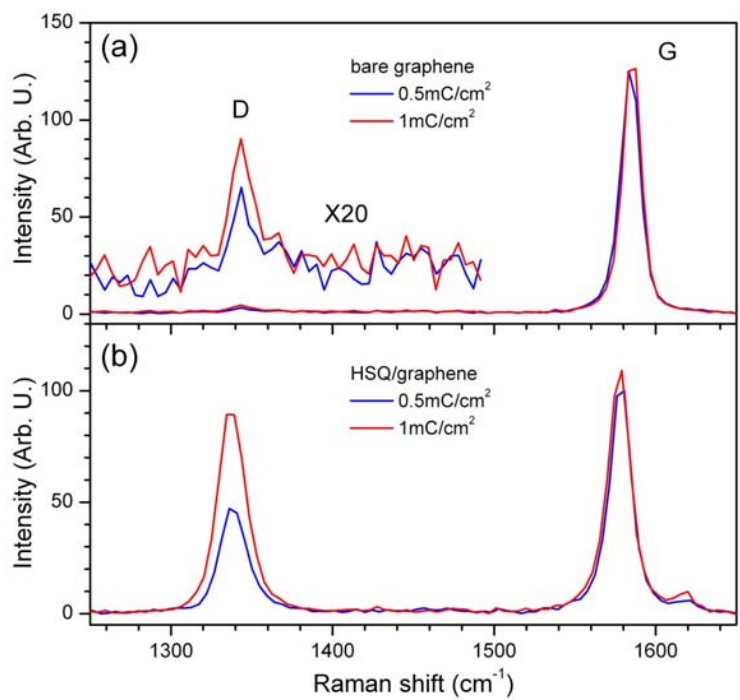

**Figure S1.** The effects of electron beam irradiation on the Raman D band of 1L graphene taken in ambient conditions: (a) bare graphene (sample **IV**) and (b) HSQ-coated graphene (sample **III**). The kinetic energy of electrons was 30 keV and electron dose is given in the legend. Entire graphene sheets were uniformly irradiated with electrons for (a) and (b). For the Raman measurements, $\lambda_{exc}$ = 514.5 nm.

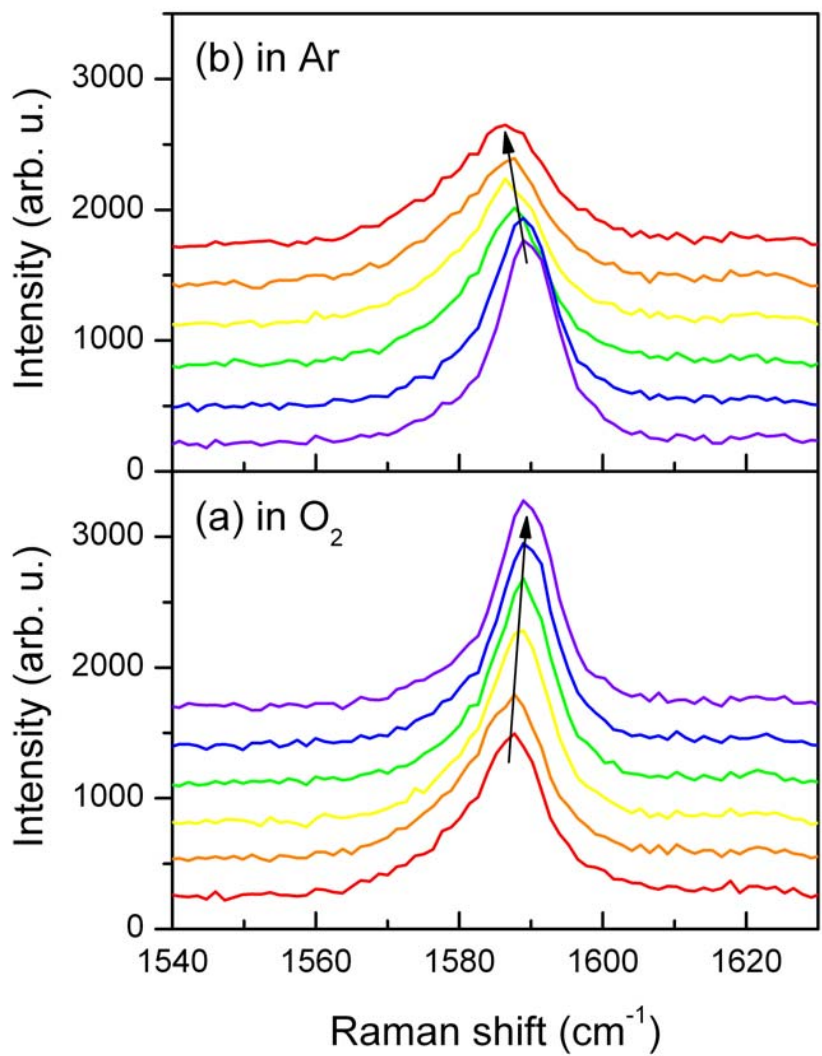

**Figure S2.** The effect of ambient gas on the G band of 1L graphene (sample **I**) dehydrogenated by oven annealing at 100 °C in air followed by photothermal heating as described in the Supporting Information text. (a) A few hour-long prior Ar flow through the gas flow cell was replaced by $O_2$ flow at the beginning of the first measurement (bottom spectrum). (b) Prior $O_2$ flow was replaced by Ar flow at the beginning of the first measurement (bottom spectrum). The individual spectra in (a) and (b) (displaced for clarity) were taken consecutively at 5 minute intervals in time. The arrows trace the evolution of the Raman G band energy with elapsed time. For the Raman measurements, $\lambda_{exc}$ = 632.8 nm.

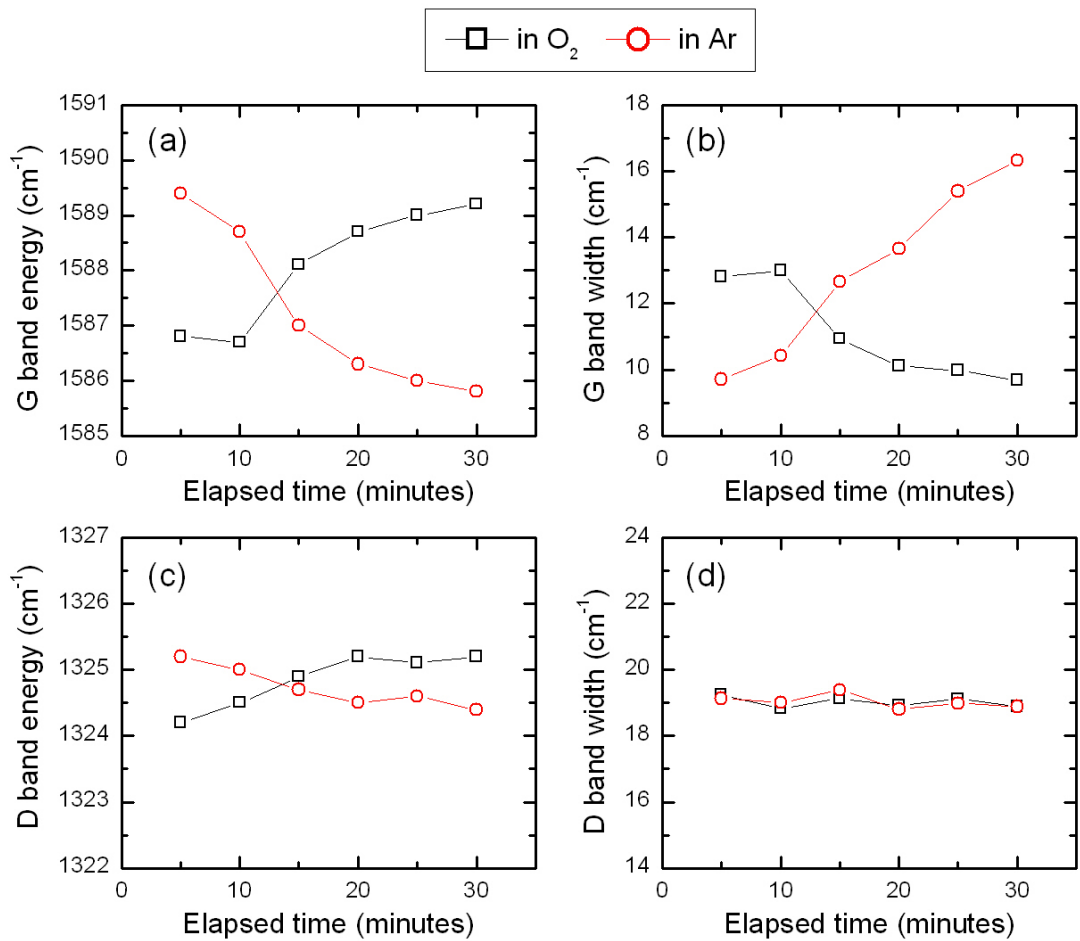

**Figure S3.** The effect of ambient gas on the G (a, b) and D (c, d) bands of 1L graphene (sample **I**) dehydrogenated at 100 °C. The data were obtained by fitting the spectra in Fig. S2. The line width includes an instrument response function of 3.7 cm$^{-1}$.

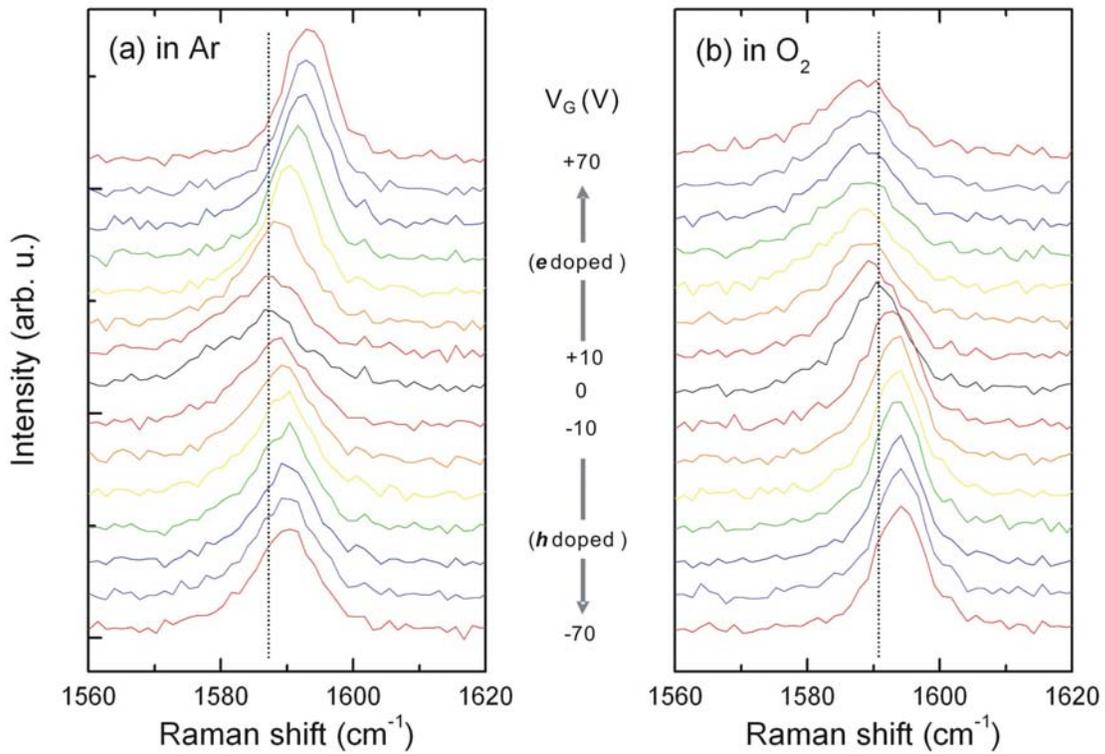

**Figure S4.** The G band Raman spectra of 1L graphene (sample **I**), dehydrogenated at 100 °C, obtained as a function of back gate voltage ($V_G$) in Ar and $O_2$. The spectra (displaced for clarity) correspond to steps of 10 V in $V_G$. The wavelength of the Raman pump laser was $\lambda_{exc} = 632.8$ nm and the irradiance was approximately 130 kW/cm$^2$.